\documentclass[prl,twocolumn, a4paper, superscriptaddress]{revtex4-1}
\usepackage[utf8]{inputenc}
\usepackage[pdftex]{graphicx}
\usepackage{color}
\usepackage{amsmath}
\usepackage{amssymb}

\newcommand{\blue}[1]{\textcolor{black}{#1}}

\usepackage{mathtools}
\DeclarePairedDelimiter{\evdel}{\langle}{\rangle}

\begin{document}

\title{Adhesion-induced Discontinuous Transitions \\and Classifying Social Networks}

\author{Nora Molkenthin}
 \affiliation{Network Dynamics, Max Planck Institute for Dynamics and Self-Organization (MPIDS), 37077 Göttingen, Germany}
\author{Malte Schröder}
 \affiliation{Chair for Network Dynamics, Center for Advancing Electronics Dresden (cfaed) and Institute for Theoretical Physics, Technical University of Dresden, 01069 Dresden}
 \affiliation{Network Dynamics, Max Planck Institute for Dynamics and Self-Organization (MPIDS), 37077 Göttingen, Germany}
 \author{Marc Timme}
 \affiliation{Chair for Network Dynamics, Center for Advancing Electronics Dresden (cfaed) and Institute for Theoretical Physics, Technical University of Dresden, 01069 Dresden}
 \affiliation{Network Dynamics, Max Planck Institute for Dynamics and Self-Organization (MPIDS), 37077 Göttingen, Germany}
\begin{abstract}
Transition points mark qualitative changes in the macroscopic properties of large complex systems. Explosive transitions, exhibiting properties of both continuous and discontinuous phase transitions, have recently been uncovered in network growth processes. Real networks not only grow but often also restructure, yet common network restructuring processes, such as small world rewiring, do not exhibit phase transitions. Here, we uncover a class of intrinsically discontinuous transitions emerging in network restructuring processes controlled by \emph{adhesion} --  the preference of a chosen link to remain connected to its end node. Deriving a master equation for the temporal network evolution and working out an analytic solution, we identify genuinely discontinuous transitions in non-growing networks, separating qualitatively distinct phases with monotonic and with peaked degree distributions. Intriguingly, our analysis of heuristic data indicates a separation between the same two forms of degree distributions distinguishing abstract from face-to-face social networks.
\end{abstract}
\maketitle
Phase transitions mark qualitative changes in the collective behavior of large complex systems by separating regimes of distinct collective states.  
The change of the global system state is induced by a control parameter passing a critical value.
Common examples include the liquid-gas transition, where the fluid density drops discontinuously upon smoothly decreasing the pressure 
and the transition between magnetic and non-magnetic states in ferromagnetic solids with increasing temperature.

Recent studies  have revealed new forms of `explosive' transitions in the macroscopic structure of networks \cite{Achlioptas2009,Bohman2009,Stauffer1994a,Cohen2010}.
The network ensembles in these examples originate from a growth process, where links are added sequentially.
While initially deemed discontinuous, these transitions have since been shown to be more subtle
\cite{Nagler2011,Ziff2009,Schroder2013,Boccaletti2016,Fronczak}. 
Despite strong analogies to finite size and super-critical properties of discontinuous phase transitions \cite{Nagler2011,DSouza2015} 
they are in fact
continuous \cite{Riordan2011,DaCosta2010}. 
Some notable exceptions have been found under specific conditions, for instance for two-layer networks,
globally optimized link selection or inhomogeneous node preferences \cite{Chen2015,DSouza2014,bassler2015extreme}. 
Studies of models of non-growing networks with implicitly degree-dependent rewiring  \cite{lindquist2009network} found different types of degree distributions. The question about existence and type of phase transitions has not been addressed.
At the same time, a range of processes that restructure rather than grow the system are known to \textit{not} exhibit phase transitions but instead show a gradual crossover between the different network structures \cite{Watts1998,grosskinsky2002universal,grabow2012small,watts1999small,newman2000mean,Molkenthin2016a}.
Under which conditions phase transitions may possibly emerge in non-growing systems and whether they could be discontinuous thus remains unknown to date.

\begin{figure}[t]
        \centering
        \includegraphics[width=\columnwidth]{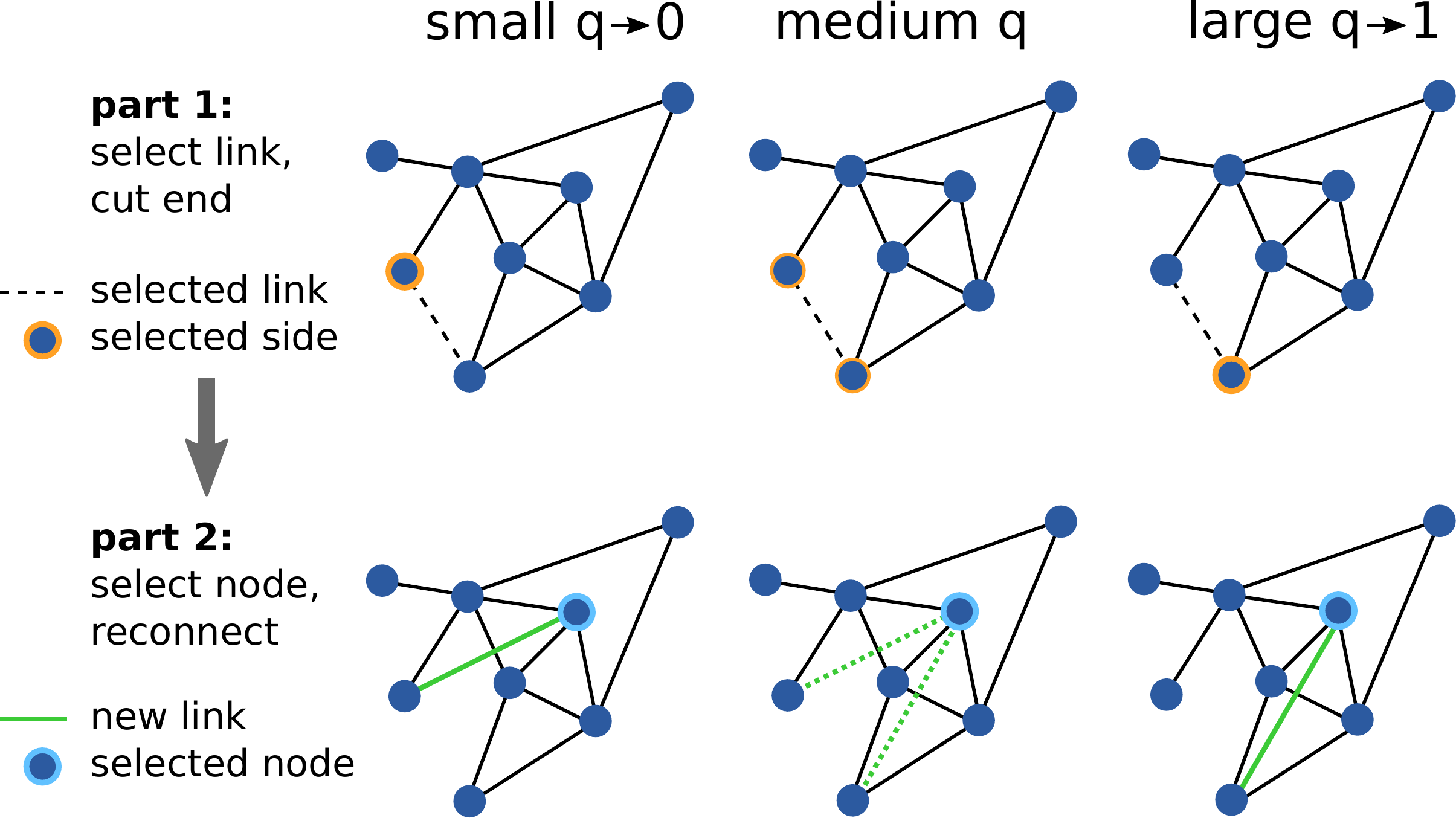}
        \caption{\textbf{Network rewiring controlled by link adhesion.}
        Rewiring at each time step happens in two subsequent parts. First, a link is randomly selected and one of its ends is cut according to its adhesion preference $q$, setting the probability for cutting the chosen link's lower degree end. As $q\rightarrow 0$, the higher degree end is cut preferentially, as $q\rightarrow 1$, the lower-degree side is cut preferentially and if $q = 0.5$ both sides are cut with the same probability. Second, the cut end of the link is reconnected to a node of degree $k$ with probability proportional to $k+1$ (preferential attachment). 
        \label{fig.process}
        }
\end{figure}

\begin{figure}[t]
        \centering
        \includegraphics[width=\columnwidth]{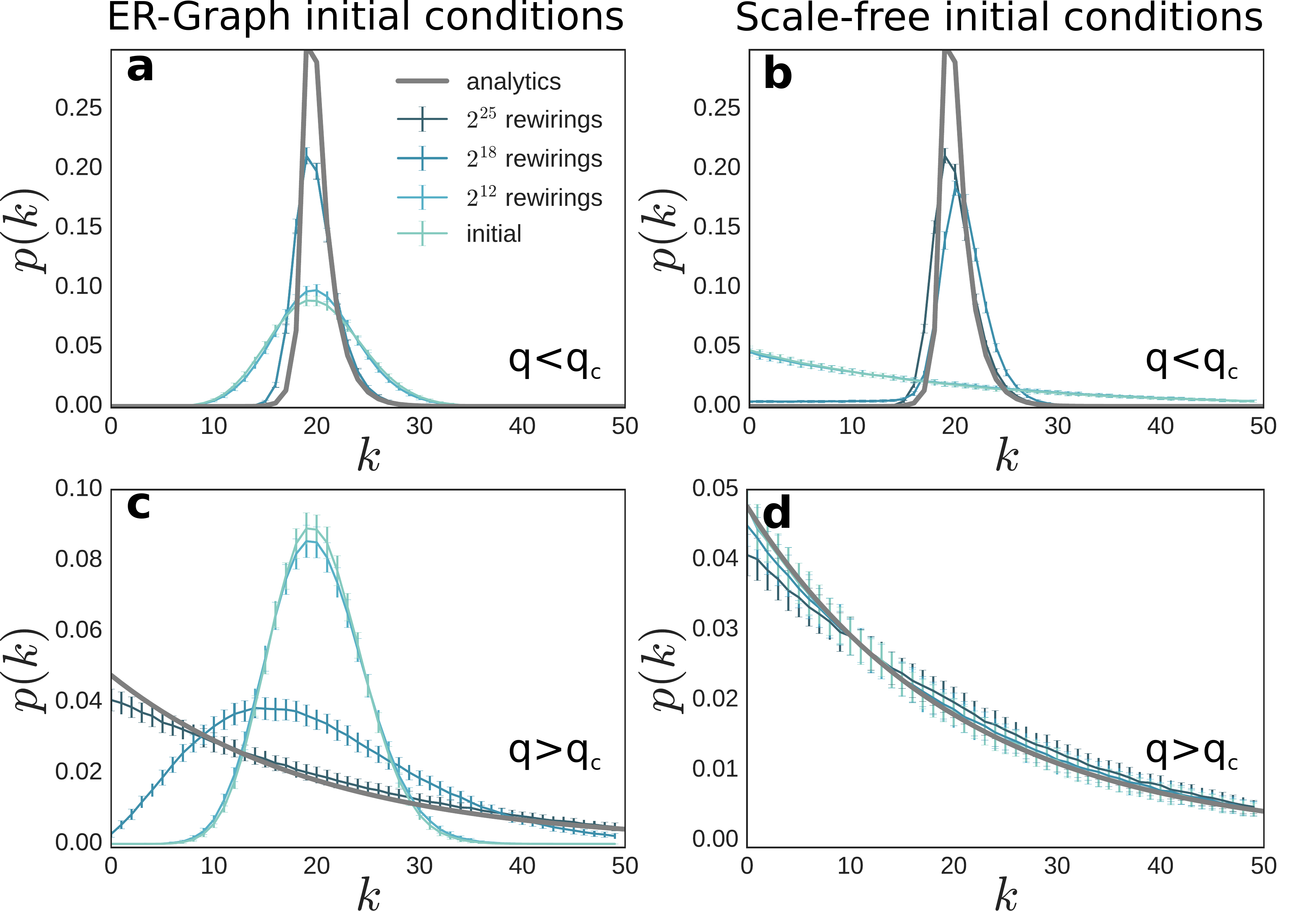}
        \caption{
        \textbf{Qualitatively different steady-state degree distributions for different $q$.} 
        a), b) For small $q$ ($q=0.1$ in the examples) the stationary degree distributions are peaked at $k=\left<k\right>-1 $ whereas c), d), for larger $q$ ($q=0.5$ in the example), they become monotonically decreasing in $k$, thus peaking for $k=0$. The transition is discontinuous (jumping from peak position $\left<k\right>$ to zero) and robust against changes of initial conditions (Erdős-Rényi initial networks, panels a and c; Barabási-Albert initial networks, panels b and d). Example ensembles shown for initial networks of $N=2^{12}$ nodes and $\left< k\right>=20$. Darker colors indicate longer times and gray wide lines indicate the analytical prediction as time $t\rightarrow \infty$ (stationary distribution). Data are averaged across 100 realizations of the stochastic temporal network evolution; error bars show standard deviation of resulting distribution. Note the different vertical scales in panel d) compared to c).}
 \label{fig.sim}
\end{figure}

Here, we study a class of network restructuring processes with explicit \emph{adhesion} preference of a link to stay connected to its end nodes. The resulting networks exhibit a genuine discontinuous phase transition in their macroscopic structure as an adhesion parameter changes.
A degree-dependent asymmetry of the node-link adhesion
induces a transition between classes of networks with qualitatively different degree distributions, one monotonically decaying with degree, the other peaked. 
Intriguingly, the two  classes consistently distinguish abstract (e.g. online) from face-to-face (e.g. offline) social networks as we confirm by comparing analytical and simulational results of our theory with structural network data for empirical systems. 

\textit{Degree-dependent network restructuring.} Consider a basic restructuring process of a non-growing temporal network of $N$ nodes and $L$ links, starting from an arbitrary interaction topology. In each time step
$t\in\mathbb{N}_0$ of restructuring (Fig.\,\ref{fig.process}), a link in the network is chosen at random from the uniform distribution among all links.
One of the two end nodes of that link is cut (`given up' by the unit it connected to), with probability $q$ choosing the lower-degree node and with probability $1-q$ the higher-degree node. The cut end of the link reconnects to a different unit of some degree $k$, randomly chosen with a probability distribution proportional to $(k+1)$.

The resulting time evolution defines a stochastic ensemble of temporal networks. After sufficiently long times, we consistently observe convergence to a network ensemble dependent only on $q$ (unique attractor). Indeed, the degree distributions become stationary and independent of the initial network structure (Fig.\,\ref{fig.sim}). For low $q$ the stationary distribution is peaked, for higher $q$ it is decaying.
Below, we evaluate
the order parameter
\begin{equation}
m = \evdel{k}^{-1} \mathrm{argmax}[\tilde{P}(k)] 
\label{eq.mk}
\end{equation}
given by the normalized mode (position of the maximum) of the stationary degree distribution $\tilde{P}(k)$ and show that the system exhibits a well-defined phase transition in the thermodynamic limit where $\evdel{k} \rightarrow \infty$ and thus  $N \rightarrow \infty$.

To qualitatively understand how the transition emerges and how it depends on adhesion and network topology, we derive and analyze a master equation characterizing the degree distribution $P_t(k)$ of the evolving network ensemble. 
Consider the possible degree changes at a given time step $t\geq 0$. The one end node that sticks at the randomly chosen link does not change its degree, whereas the other end node reduces its degree by one. Subsequently, the node that the link rewires to increases its degree by one. As a consequence, the degree distribution of the temporal network process satisfies the discrete-time master equation
\begin{align}
P_{t+1}(k)&=P_t(k)-u_k P_t(k)+u_{k+1}P_t(k+1)\nonumber \\
&-l_k P_t(k)+l_{k-1}P_t(k-1).
\label{eq.rec2}
\end{align}
The unwiring probability
$u_k$ represents the probability of decreasing the degree of one of the nodes with degree $k$ by rewiring away from it and 
the linking probability $l_k$ the probability of increasing the degree of a node with degree $k$ by rewiring to it. 

As the temporal network evolution is a Markov chain that is irreducible, i.e. every network in the ensemble can be reached with positive probability, the stationary degree distribution ${\tilde{P}}(k)$
is unique and given as solution of Eq.\,\ref{eq.rec2} with $P_{t+1}(k)=P_t(k)=\tilde{P}(k)$.
We rewrite Eq.\,\ref{eq.rec2} at the fixed point as a matrix equation 
\begin{equation}
 \begin{aligned}
\underbrace{
  \begin{pmatrix}
  \text{-}(l_0) & u_1 & 0 &\cdots & 0\\
  l_0 & \text{-}(l_1+u_1) & u_2 &\cdots & 0 \\
  \vdots &\vdots & \ddots &\vdots& \vdots  \\
  0 & \cdots&0 & l_{N\text{-}2} & \text{-}u_{N\text{-}1} 
 \end{pmatrix}}_{\mathbf{M}}
    \begin{pmatrix}
  {\tilde{P}}(0)\\
  {\tilde{P}}(1) \\
  \vdots \\
  {\tilde{P}}(N\text{-}1)
 \end{pmatrix}=0.
\end{aligned} 
\label{eq.fix2}
\end{equation}
where ${\tilde{P}}$ is a vector with entries ${\tilde{P}}(k)$ and $\mathbf{M}$ a matrix with entries based on $u_k$ and $l_k$.
Adding to each row of the matrix $\mathbf{M}$ the previous row and dividing by $u_i$ simplifies Eq.\,\ref{eq.fix2} to 
\begin{align}
\underbrace{
 \begin{pmatrix}
  \frac{l_{0}}{u_1} & -1 & 0 &\cdots & 0\\
  0 & \frac{l_{1}}{u_2} & -1 &\cdots & 0 \\
  \vdots &\vdots & \ddots &\vdots& \vdots  \\
  0 & \cdots&0 & 0 & 0
 \end{pmatrix}}_{\mathbf{M}'}
\begin{pmatrix}
  {\tilde{P}}(0)\\
  {\tilde{P}}(1) \\
  \vdots \\
  {\tilde{P}}(N-1)
 \end{pmatrix}=0.
\label{eq.fix3}
\end{align}

Since the last row of $\mathbf{M}'$ is identically zero, $\mathbf{M}'$ does not have full rank and there exists a stationary solution 
for $k \in \{1,\ldots,N-1\}$,
given by the eigenvector to the eigenvalue zero as
\begin{align}
p(k)&=
\left(\prod_{m=1}^{k}\frac{l_{m-1}}{u_m}\right).
 \label{eq.Pgen}
\end{align}
We fix $p(0)$ arbitrarily and then normalize $\tilde{P}(k)=Cp(k)$ by $C=1/\sum_{k=0}^{N-1} {p}(k)$ for all k such that $\sum_{k=0}^{N-1} {\tilde{P}}(k)=1$
to obtain the exact stationary degree distribution.
\begin{figure}[!th]
        \centering
        \includegraphics[width=\columnwidth]{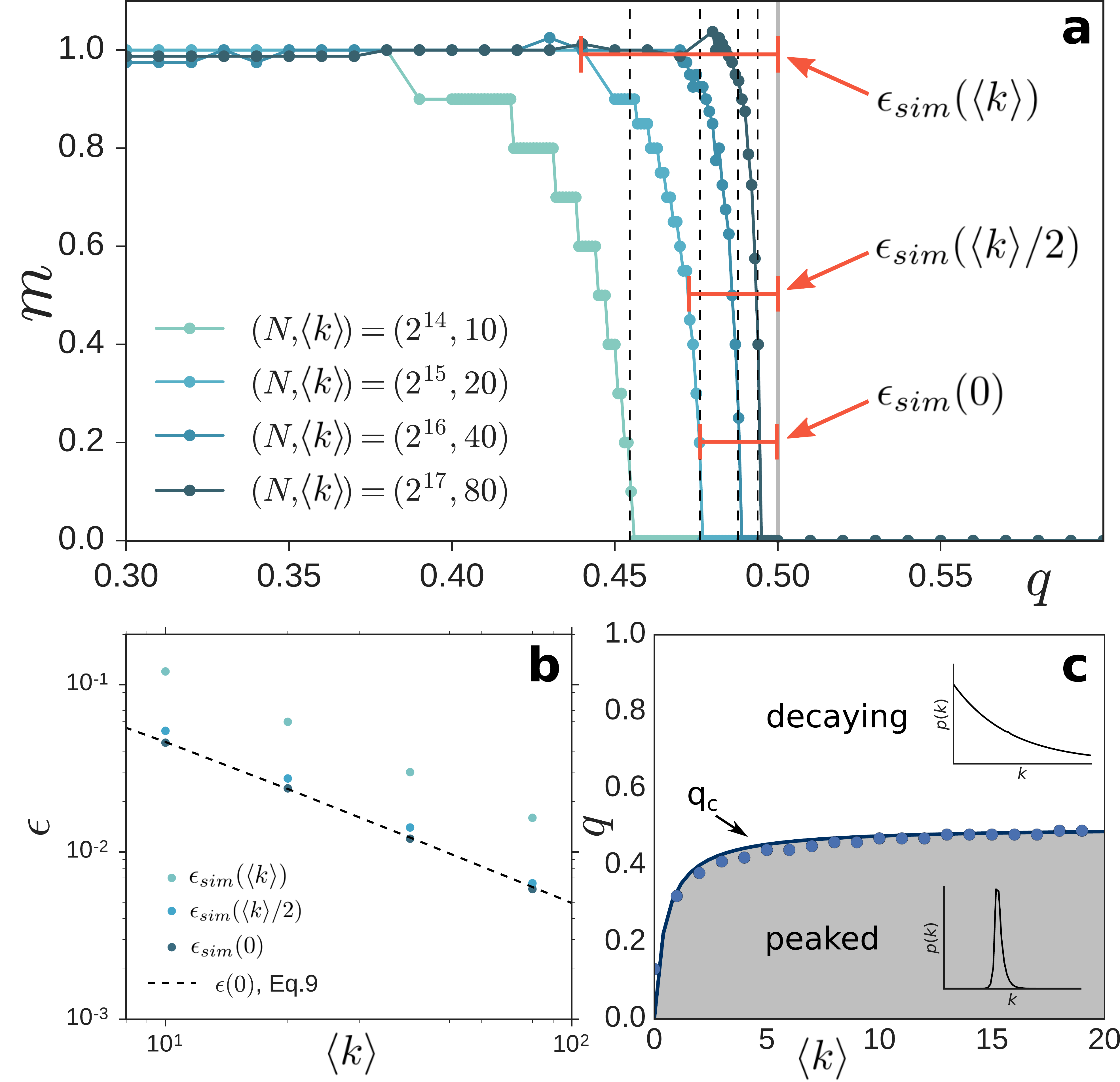}
        \caption{\textbf{Phase transition in temporal network rewiring}. 
        a) The relative position of the maximum $m =\evdel{k}^{-1} \mathrm{argmax}[\tilde{P}(k)]$ of the stationary degree distribution $\tilde{P}(k)$ as a function of adhesion asymmetry parameter $q$ for different network sizes with $N > \evdel{k} \gg 1$, where $\evdel{k}/N = \mathrm{const.}$
        b) The distance $\epsilon(k)$ from the transition point (\ref{eq.epsilon})  algebraically decays to zero in the thermodynamic limit for all $k\leq \evdel{k}$, see the examples for $k/\evdel{k}=\{0,1/2,1\}$, thus indicating a discontinuous transition at $q_c = 1/2$. The scaling $\epsilon(k) \sim \evdel{k}^{-1}$ agrees with the analytical approximation  (\ref{eq.eps}) and the exact prediction (\ref{eq.eps0}).
        c) Phase diagram in parameter space $q$ vs. $\evdel{k}$ indicates separation between decaying ($m = 0$) and peaked ($m > 0$) degree distributions (white/shaded regions, resp.). The analytical prediction (solid line) shows $q_c(0)$ from Eqns.~(\ref{eq.eps0}) and (\ref{eq.epsilon}), disks show the result from direct numerical simulations for $N = 1024$.}
        \label{fig.pt}
\end{figure}

\textit{Asymmetric adhesion induces discontinuous transitions.} To calculate the peak position 
$k_\mathrm{max}=\mathrm{argmax}[\tilde{P}(k)]$ of the degree distribution, we first consider the unlinking and linking probabilities
$u_k$ and $l_k$.  Since links are selected uniformly at random, the probability of a link with an end node of degree $k$ being selected is $k/L$, where $L = N\evdel{k}/2$ is the number of links in the network. That node is chosen to be cut from the link with probability $q$ if $k$ is smaller than the degree of the other end node of the link and with probability $1-q$ if it is larger than that degree. 
For uncorrelated degrees of the two end nodes we obtain the approximation
\begin{align}
u_k&=\frac{k}{L}\left(q P^{\scriptscriptstyle>}(k)+ \frac{1}{2} P^{\scriptscriptstyle=}(k) + (1-q) P^{\scriptscriptstyle<}(k)\right),
\label{eq.pu}
\end{align}
where $P^{\scriptscriptstyle>}(k) = \sum_{k'=k+1}^{N-1} \frac{k'}{\evdel{k}}\tilde{P}(k')$, $P^{\scriptscriptstyle=}(k) = \frac{k}{\evdel{k}}\tilde{P}(k)$ and $P^{\scriptscriptstyle<}(k)=\sum_{k'=1}^{k-1}\frac{k'}{\evdel{k}}\tilde{P}(k')$ are the probabilities that a node neighbouring a degree $k$ node has itself a degree larger than, equal to, or smaller than $k$. Numerical results suggest that for networks where $\evdel{k}$ is substantially smaller than $N$ the node degrees are indeed sufficiently weakly correlated in the stationary state for $q \le 1/2$.

The cut link rewires following preferential attachment \cite{Barabasi1999} and rewires to a new node with probability proportional to $(k+1)$.
The offset of $1$ prevents a node from being removed if its degree falls to zero.
So the (linking) probability of reconnecting to a node of degree $k$ is 
\begin{equation}
l_k = \frac{k+1}{\sum_{i=1}^{N}(k_i+1)}=\frac{k+1}{N\evdel{k+1}}.
\label{eq.pl}
\end{equation}
\begin{figure*}[t]
        \centering
        \includegraphics[width=16 cm]{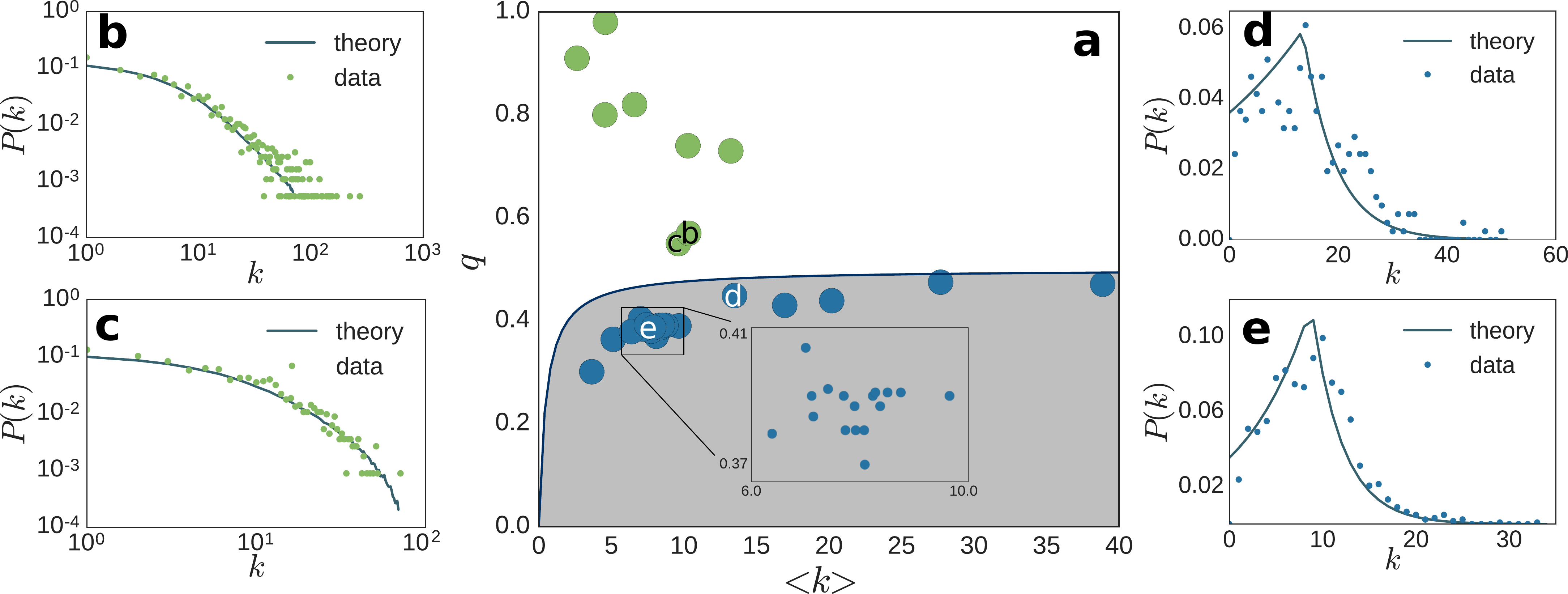}
        \caption{\textbf{Social networks in the phase diagram.} a) The phase transition (dark blue line) theoretically derived from the simple network restucturing model separates all personal (dark blue) from all abstract (light green) social networks studied. The topologies of 33 different networks were extracted from data collected in \cite{Kunegis2013}. For each network $i$, the optimal parameter $q_i$ given the average degree $\evdel{k}_i$ (and network size $N_i$) as extracted from the network data, are placed in the phase diagram. The inset magnifies a densely covered area of high school social networks.
        b-e) Examples of social networks illustrating decaying and peaked degree distributions, the solid curves indicate single-parameter fit of the entire distributions. b) Online community ``Hamsterster'' \cite{konect:2016:petster-friendships-hamster}, c) University Rovina email exchange \cite{konect:guimera03}, d) interactions in an exhibition \cite{konect:sociopatterns}, e) and friendships among school children \cite{konect:moody}.}
        \label{fig.data}
\end{figure*}
To understand how the peak position changes, we calculate the values of $q=:q_c(k)$ where the maximum of the degree distribution changes from $k$ to $k+1$ in the steady state ensemble. We then demonstrate that $q_c(k) \rightarrow 1/2$ for all $k\leq \evdel{k}$ in the limit as $\evdel{k}\rightarrow \infty$ and thus $N\rightarrow \infty$  such that the order parameter discontinuously jumps at that value of $q$. Fig.~\ref{fig.pt} displays finite system results and illustrates the scaling as $\evdel{k}$ and $N$ grow. For finite systems, the point $q_c(k)$ is defined by the condition that the two consecutive probabilities become equal, $\tilde{P}(k) = \tilde{P}(k+1)$.

Using $P^{\scriptscriptstyle>}(k) + P^{\scriptscriptstyle=}(k) + P^{\scriptscriptstyle<}(k) = 1$ and writing the probabilities $P^{\scriptscriptstyle>}(k) = (1 - P^{\scriptscriptstyle=}(k))/2 + \delta^{\scriptscriptstyle>}(k)$ and $P^{\scriptscriptstyle<}(k) = (1 - P^{\scriptscriptstyle=}(k))/2 + \delta^{\scriptscriptstyle<}(k)$, where $\delta^{\scriptscriptstyle>}(k) = - \delta^{\scriptscriptstyle<}(k)$, 
 Eqs.~(\ref{eq.Pgen} - \ref{eq.pl}) yield the equation
\begin{align}
	1 &= \frac{\tilde{P}(k)}{\tilde{P}(k+1)} = \frac{u_{k+1}}{l_k} \nonumber \\
	&=\frac{2\evdel{k+1}}{\evdel{k}} \left[ \frac{1}{2} + \left(2q_c(k)-1\right) \delta^{\scriptscriptstyle>}(k) \right] \label{eq.delta}
\end{align}
for $q_c(k)$. For $q_c = 1/2$ the right-hand-side is larger than one for all finite $\evdel{k}$. 
We thus consider the deviation
\begin{align}
\epsilon(k) = 1/2 - q_c(k)  \geq 0 \,, \label{eq.epsilon}
\end{align}
such that (\ref{eq.delta}) results in 
\begin{align}
	\epsilon(k) &= \frac{1}{4 \evdel{k+1} \delta^{\scriptscriptstyle>}(k)} \,. \label{eq.eps}
\end{align}
This provides strong evidence for a discontinuous phase transition, because in the limit $\evdel{k} \rightarrow \infty$ we have $\epsilon(k)\rightarrow 0$ for all $k$ as long as $\delta^{\scriptscriptstyle>}(k) > 0$. We make additional progress by noting that $\delta^{\scriptscriptstyle>}(k)$ changes sign at the median $\bar{k}$ of the distribution. For the typical (though not universal) ordering of the mode $k_\mathrm{max}$, median $\bar{k}$ and mean $\evdel{k}$ of a unimodal distribution with positive skewness \cite{groeneveld1977mode,abadir2005mean}, $k_\mathrm{max} \le \bar{k} \le \evdel{k}$, we thus find that indeed $\delta^{\scriptscriptstyle>}(k_\mathrm{max}) > 0$ and $q_c(k) < 1/2$, consistent with the transitions observed in the simulations of finite systems (Fig.~\ref{fig.pt}a). As a result, the peak position changes discontinuously from $m=\frac{k_\mathrm{max}}{\evdel{k}} = 0$ to $m=1$ at $q_c = 1/2$ in the limit of $\evdel{k} \rightarrow \infty$ and thus $N\rightarrow \infty$.

For $k=0$ we specifically have $P^{\scriptscriptstyle>}(0) = 1$ and $P^{\scriptscriptstyle=}(0) = 0$ and consequently $\delta^{\scriptscriptstyle>}(0) = 1/2$ such that we obtain the exact expression
\begin{align}
	\epsilon(0) &= \frac{1}{2 \evdel{k+1}} \,. \label{eq.eps0}
\end{align}
Direct numerical simulations show that the same scaling holds for $\epsilon(k)$ across values of $k \in \{0,...,\evdel{k}\}$, compare Fig.~\ref{fig.pt}b

The approximate scaling form (\ref{eq.eps}), the exact result (\ref{eq.eps0}) and the numerical analysis summarized in Fig.~\ref{fig.pt} jointly indicate a discontinuous phase transition at $q_c=1/2$ in the limit $\evdel{k} \rightarrow \infty$, and thus $N\rightarrow \infty$. We emphasize that the order parameter changes abruptly from $m = 1$ for all $q<1/2$ to $m = 0$ for $q>1/2$ as the adhesion parameter $q$ continuously varies across it's critical value from below.

\textit{Transition line separates abstract from personal social networks.} Interestingly, a wide range of social networks exhibit one of these two specific types of degree distributions -- decaying or peaked -- theoretically identified above. Moreover, the theoretical transition line exactly separates the collection of networks into those with
close personal contacts and those created through more abstract, indirect or online-only interactions, compare Fig.\,\ref{fig.data}: 

We have compared the degree distributions resulting from the simple model 
to those obtained from 33 social networks, as reported in references \cite{Kunegis2013} and \cite{starbuck,konect:boguna,konect:dolphins,konect:coleman1957,konect:ucidata-zachary,konect:arenas-jazz,konect:freeman1998,konect:knuth1993,konect:boguna,DBLP:Massa}.
The systems range from networks of direct personal interactions with face-to-face contacts in various private and educational contexts or reported friendships of humans and of dolphins, to more abstract social interactions, including online social networks as well as offline, but hierarchically 
determined social relations (see supplemental material for more details).
The average degree $\left<k\right>$ was computed from each data set thus leaving $q$ as the only free parameter. A least-squares fit to each degree distribution yields the best-suitable $q$ for each network. 
Intriguingly this even led to good quantitative agreement of the degree distributions (see Fig.\,\ref{fig.data}\,b-e, \blue{see also supplemental material}).

We were amazed to observe that all networks in which personal, typically face-to-face relationships define links exhibit a peaked degree distribution, whereas all networks in which abstract, i.e. online relationships define links, exhibit decaying degree distributions, see Fig.\,\ref{fig.data}. Indeed, a permutation test  assigning the labels 'abstract' and 'personal' to the degree distributions randomly yields the same classification only in one out of $\binom{33}{8}\approx 1.4\times 10^7$ cases.

\textit{Discussion and Conclusions.} Studies of network structure forming processes have previously uncovered explosive transitions in a range of network growth processes \cite{DSouza2015} whereas most restructuring processes for fixed size networks
exhibit gradual cross-overs between random and regular graphs with no distinct transition. Notably, the rewiring mechanisms in the latter types of processes are independent of any properties of the nodes or links \cite{Watts1998, grosskinsky2002universal,grabow2012small,watts1999small,newman2000mean,Molkenthin2016a}. 

Here, we revealed a discontinuous phase transition in network structural features in a simple model class of temporal networks that do not grow but exhibit degree-dependent link adhesion. As the underlying microscopic mechanisms rely on a simple local cutting and rewiring process, they imply self-organization of the networks' large-scale structures.  Driven by the tendency of a link to stick to a node (adhesion) depending on that node's degree, a discontinuous phase  transition emerges between two types of degree distributions -- peaked and decaying -- when smoothly varying this tendency via a control parameter. 

Intriguingly, the same two types of degree distributions are observed for a wide range of social networks. Moreover, among the 33 topologies of social networks analyzed, all those networks established through personal contacts exhibit peaked and all those with more abstract, indirect or impersonal contacts exhibit decaying degree distributions. The results thus suggest that the simple, abstract model, that is \emph{a priori} unrelated to specific and widely heterogeneous social dynamics, surprisingly serves as a good indicator for the separation between networks with more personal and more abstract social relations.

Several general constraints as well as social mechanisms might be supporting this binary classification. The primary mechanisms underlying the model restructuring process are based on the model ingredient that sustaining a link is node-dependent. From the perspective of a node in a socio-economic setting, such as trade or friendships, this may be interpreted as the 
bilateral effort (e.g., time, money or motivation) to keep a connection. 
Such constraints on the number of sustainable links may lead a node to cut ties with less beneficial connections (i.e. a less involved friend or a less influential business partner), similar to our model settings that otherwise are far from fully capturing the intricacies across social networks.

Taken together, the results presented above not only highlight severe theoretical consequences of link adhesion -- inducing a discontinuous transition for restructuring networks in the first place -- and yield novel insights into  structural phase transitions in temporally evolving networks \cite{HOLME201297,PhysRevLett.110.198701,scholtes2014causality}. They may also serve as a starting point for future investigations about the mechanisms and the influence of constraints in evolving socio-economic systems.

\textit{Acknowledgements.} We thank Stefan Grosskinsky, Yael Fender, Frank Schweitzer, Alex Arenas and Johanna Wolter for valuable comments on presentations about this work. This work is supported through the German Science Foundation (DFG) by a grant towards  the Cluster of Excellence `Center for Advancing Electronics Dresden' (cfaed).

\bibliographystyle{unsrt}
\bibliography{socialnets.bib}

\end{document}